\documentclass[floatfix,amssymb,prd,twocolumn,superscriptaddress,nofootinbib,preprintnumbers]{revtex4-2}

\usepackage{natbib}
\usepackage{subcaption,ragged2e}
\DeclareCaptionJustification{justified}{\justifying}
\usepackage{amssymb,amsmath,verbatim,mathtools,needspace,enumitem,etoolbox,graphicx,microtype,afterpage,bm}
\usepackage{tikz}
\usetikzlibrary{shapes.misc, positioning, arrows.meta}
\usepackage{framed}

\usepackage{hyperref}
\definecolor{linkcolor}{rgb}{0.0, 0.47, 0.75}
\definecolor{citecolor}{rgb}{1.0, 0.5, 0.0}
\hypersetup{
linkcolor = linkcolor, citecolor = linkcolor, urlcolor = linkcolor, colorlinks = true }
\usepackage{multirow,pifont,lmodern,tabularx,booktabs}
\usepackage[all]{hypcap}
\usepackage[T1]{fontenc}
\usepackage[utf8]{inputenc}
\usepackage{cleveref}

\allowdisplaybreaks
\usepackage{xcolor}

\newcommand{\IoA}{Institute of Astronomy, University of Cambridge, Madingley Road, Cambridge CB3 0HA, UK}
\newcommand{\UU}{Institute for Gravitational and Subatomic Physics, Universiteit Utrecht, Princetonplein 1, 3584 CC Utrecht, The Netherlands}
\newcommand{\Nikhef}{Nikhef, Science Park 105, 1098 XG Amsterdam, The Netherlands}
\newcommand{\Cav}{Cavendish Laboratory, University of Cambridge, JJ Thomson Avenue, Cambridge CB3 0US, UK}

\begin{document}
\title{Ab Initio Real-Time Gravitational-Wave Parameter Estimation}
\date{\today}

\author{David~Yallup}
\email{dy297@cam.ac.uk}
\affiliation{\IoA}

\author{Metha~Prathaban}
\affiliation{\Cav}

\author{James~Alvey}
\affiliation{\IoA}

\author{Thomas~C.~K.~Ng}
\affiliation{\Nikhef}
\affiliation{\UU}

\author{Thibeau~Wouters}
\affiliation{\UU}
\affiliation{\Nikhef}

\author{Nikhil~Sarin}
\affiliation{\IoA}

\author{Will~Handley}
\affiliation{\IoA}

\begin{abstract}
\noindent
We present a specialised GPU-native nested sampling kernel targeting rapid parameter estimation for gravitational wave inference problems. Building upon a Slice-within-Gibbs (SwiG) structure for rapid mixing, we investigate how far we can push baseline stochastic sampling techniques on modern GPU hardware. We demonstrate that for typical long-duration binary neutron star signals observed by the LIGO and Virgo detectors, we can achieve well calibrated posterior inference on the full uncompressed data of a three detector network in a median of twelve minutes on a single GPU. This falls to five minutes when sharded across four devices. Utilising heterodyning to compress the data reduces the median wall time across an injection campaign to 89 seconds -- less than the length of the segment itself -- and enables inference with precessing spin, tidal waveforms on GW170817 in around two minutes. This pushes stochastic sampling techniques using full physical waveform calculations, launched from an uninformed prior state, towards real-time gravitational wave parameter estimation.
\end{abstract}
\maketitle

\preprint{}

\section{Introduction}\label{sec:intro}

\noindent Since the first direct detection~\citep{LIGOScientific:2016aoc}, $390$ gravitational waves (GWs) emitted by mergers between black holes (BHs) and neutron stars (NSs) have been confidently detected~\citep{LIGOScientific:2026wfs}.
Binary neutron star (BNS) and neutron star-black hole (NSBH) mergers can give rise to electromagnetic (EM) counterparts, allowing for multi-messenger observations.
The gold standard detection of this type was GW170817~\citep{LIGOScientific:2017vwq}, where the rapid localisation within the detector network enabled full multi-messenger follow-up~\citep{LIGOScientific:2017ync}. Indeed, the global co-ordination of telescope consortia led to the detection of the accompanying short gamma-ray burst GRB~170817A~\citep{LIGOScientific:2017zic,Goldstein:2017mmi} and the detailed characterisation of the ensuing kilonova AT~2017gfo~\citep{Coulter:2017wya}.
This single event provided deep insight into the equation of state of neutron stars~\citep{LIGOScientific:2018hze}, the propagation speed of gravitational waves~\citep{LIGOScientific:2017zic} and the expansion rate of the Universe~\citep{LIGOScientific:2017adf}.
So far, GW170817 has been the only multi-messenger observation of this kind. For other BNS or NSBH mergers, no confident EM counterpart was detected~\citep{Pillas:2025pfc,Wouters:2025ull}. 

Future observing runs, with upgrades to existing facilities and the commissioning of more sensitive GW detectors, hold the promise to advance these science cases further by enabling more multi-messenger detections~\citep{Kiendrebeogo:2023hzf,Shah:2023ozh,Bhattacharjee:2024wyz,Kaur:2024yag}.
However, this hinges on our ability to detect the EM counterparts of bright GW signals as soon as possible for at least two reasons. First, earlier detections can track the full evolution of the light curve.
Second, there is an intrinsic latency in requesting telescope observation time and pointing the telescopes to the correct regions in the sky.
While skymaps can be produced quickly with \texttt{BAYESTAR}~\citep{Singer:2015ema}, they get updated in alerts as soon as a posterior distribution obtained with stochastic samplers is made available~\citep{Chaudhary:2023vec}.
Therefore, the ability to run full inference pipelines at the speeds required for follow-up addresses a core bottleneck in providing full fidelity statistical information during the search for EM counterparts.

Looking ahead, the necessity of fast parameter estimation (PE) for long-duration gravitational wave signals extends well beyond these rare multi-messenger events. 
For the next generation (XG) ground-based observatories, such as the Einstein Telescope (ET)~\citep{Punturo:2010zz,ET:2019dnz,Branchesi:2023mws,ET:2025xjr} and Cosmic Explorer (CE)~\citep{Reitze:2019iox,Evans:2021gyd}, analysing BBH, NSBH, and BNS mergers will come with a high computational cost~\citep{Couvares:2021ajn,Hu:2024mvn}, mainly due to the predicted higher event rates. 
Moreover, the signals will stay in band for much longer compared to current detectors -- BNS signals for hours to days, BBH signals for minutes -- due to the proposed engineering innovations that allow these facilities to push to lower frequencies~\cite{Hild:2010id}.

\begin{figure*}
    \centering
    \includegraphics[width=\linewidth]{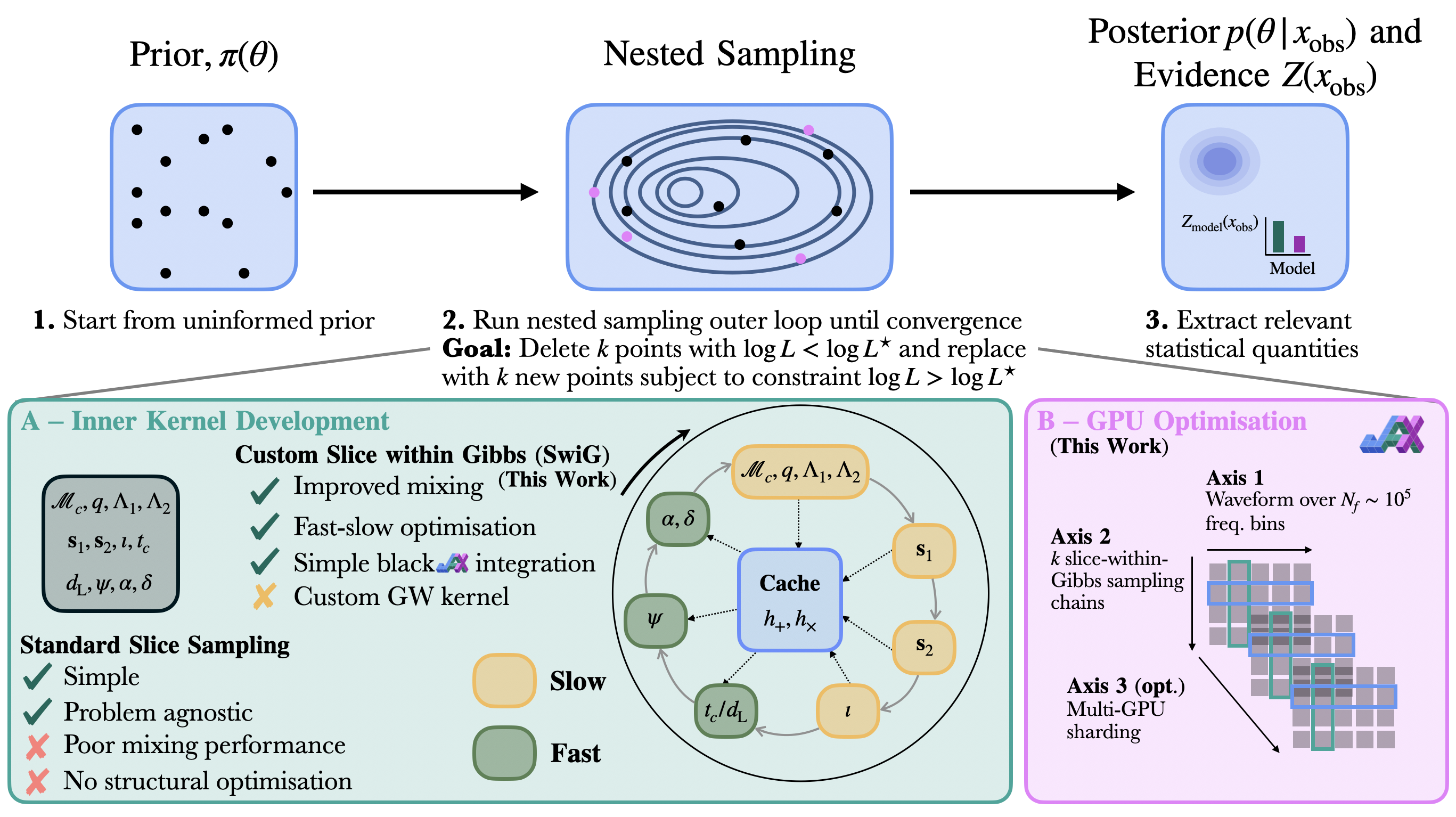}
    \caption{Schematic detailing the main contributions of this work. \textit{Top:} The conceptual flow of nested sampling, evolving a population of live points from an uninformed prior $\pi(\theta)$ over model parameters $\theta$ to the posterior $p(\theta|x_{\text{obs}})$ and evidence $\mathcal{Z}(x_{\text{obs}})$ conditioned on some observation $x_\text{obs}$. \textit{Bottom Left, Panel A:} The development of the custom Slice-within-Gibbs (SwiG) inner kernel, which improves mixing and efficiency by partitioning the parameter space into ``fast" (green) and ``slow" (yellow) blocks and utilizing waveform caching. \textit{Bottom Right, Panel B:} The three axes of GPU optimization: vectorization of the waveform across $Nf \sim 10^5$ frequency bins, parallel execution of $k$ sampling chains, and optional multi-GPU sharding.}\vspace{12pt}
    \label{fig:schematic}
\end{figure*}

The high scientific value of fast PE has therefore motivated an extensive programme of work on accelerating GW inference.
The current community-standard analyses are built on stochastic sampling. Markov chain Monte Carlo and nested sampling~\citep{Skilling2006,Ashton2022} as implemented in \texttt{LALInference}~\citep{Veitch2015} and \texttt{Bilby}~\citep{bilby_paper,Romero-Shaw:2020owr} have been established to deliver accurate, calibrated posteriors on wall times of hours to days, even when distributed across hundreds of CPU cores~\citep{2020MNRAS.498.4492S}.
Acceleration efforts, which we review in \Cref{sec:background}, have largely followed two complementary tracks: compressing the data so that each likelihood evaluation is cheaper~\citep{Cornish:2010kf,Zackay:2018qdy,Smith:2016qas,Morisaki:2021ngj}\footnote{Or, similarly, using meshfree approximations~\citep{Pathak:2022iar,Pathak:2023ixb,Sharma:2025uva}} and machine-learning methods that either assist the stochastic sampler~\citep{Williams:2021qyt,Wong:2022xvh,Wouters:2024oxj,Prathaban:2026kft} or amortise the inference entirely into a pre-trained network~\citep{Green:2020hst,Dax:2021tsq,Dax:2024mcn}.
What has so far been missing is a demonstration that the stochastic sampling paradigm itself --- exact likelihood evaluations, launched from an uninformed prior state, with no offline training stage --- can operate at the latencies that multi-messenger astronomy demands. Operating on long-duration signals also serves as a proxy to test whether algorithmic innovation within Monte Carlo methods can meet the demands of XG GW inference.

The core contribution of this work is twofold (for a schematic illustration, see \Cref{fig:schematic}). Firstly, we implement a new nested sampling algorithm, optimised for gravitational waves, based on the Slice-within-Gibbs (SwiG) kernel structure of Ref.~\cite{Yallup2026swig} and the vectorised, GPU-native nested sampling of Ref.~\cite{Yallup2026}. This kernel explicitly leverages the internal structure of compact binary coalescence (CBC) signals. Secondly, we demonstrate that when this kernel is coupled with the latest GPU hardware, we can carry out well calibrated posterior inference on the uncompressed frequency grid of a $128$\,s BNS signal in a median of twelve minutes on a single GPU, falling to five minutes when the computation is sharded across four GPUs, all with no approximation to the likelihood. Mild compression of the data brings this wall time down to $\sim 90$\,s, below the duration of the data segment itself. Finally, we validate the algorithm on GW170817, showing that had it been deployed in 2017 it would have delivered full-fidelity PE on a timescale relevant for multi-messenger follow-up. As such, this work firmly pushes stochastic sampling towards the real-time paradigm in current detectors, and illustrates a possible pathway towards large-scale XG inference.

The rest of the paper is organised as follows. \Cref{sec:background} reviews the low-latency PE landscape, nested sampling, and the SwiG kernel construction. \Cref{sec:method} instantiates this kernel for GW inference, specifying the likelihood and its analytic marginalisations, the fast-slow block structure, and the multi-GPU sharding strategy. \Cref{sec:experiments} describes the injection campaign and the sampler configurations we test, and \Cref{sec:results} presents the resulting calibration and timing studies, together with a re-analysis of GW170817. We discuss the implications and limitations of these results in \Cref{sec:discussion}, and conclude in \Cref{sec:conclusions} with an outlook towards current and next-generation GW science.

\section{Background}\label{sec:background}

\subsection{Low latency parameter estimation for gravitational waves}

\noindent Innovations to reduce the latency of GW inference have broadly come in two directions.
Firstly, one can try to downsample the number of frequency points at which waveforms are evaluated, to reduce the cost of each likelihood call.
Among such compression algorithms, the most frequently used are relative binning~\citep{Cornish:2010kf,Zackay:2018qdy,Krishna:2023bug}, reduced order quadrature (ROQ)~\citep{Canizares:2014fya,Smith:2016qas,Morisaki:2020oqk,Morisaki:2023kuq}, multibanding~\citep{Garcia-Quiros:2020qlt,Morisaki:2021ngj}, and Gaussian process interpolation~\citep{Lange:2018pyp}. 
For instance, during the past LIGO-Virgo-KAGRA (LVK)~\citep{LIGOScientific:2014pky,VIRGO:2014yos,KAGRA:2020tym} observing run, low-latency PE was achieved with ROQ acceleration, using the \texttt{IMRPhenomD}~\citep{Husa:2015iqa, Khan:2015jqa} waveform model. 
While sampling with this setup completes in less than $10$ minutes~\citep{Chaudhary:2023vec}, an analysis including spin precession and tidal deformabilities in the waveform with the \texttt{IMRPhenomPv2\_NRTidalv2} waveform~\citep{Hannam:2013oca,Khan:2015jqa,Dietrich:2019kaq} takes well over an hour~\citep{Morisaki:2023kuq}. 
Moreover, ROQ likelihoods require additional effort in building the bases. 
Finally, it has been argued that ROQ can produce fast inference for next-generation BNS signals~\citep{Smith:2016qas}, although it requires a significant memory footprint~\citep{Baker:2025taj} and correspondingly, additional techniques have to be developed to combat these memory requirements~\citep{Guttman:2026fpc}.

The other, complementary direction is to use machine learning techniques or advanced MCMC methods, either to accelerate stochastic samplers~\citep{Williams:2021qyt,Gabrie:2021tlu,Karamanis:2022alw,Karamanis:2022ksp,Williams:2023ppp,Wong:2023lgb,Wouters:2024oxj,Prathaban:2024rmu,Negri:2025cyc,Prathaban:2026kft,Williams:2025aar,Demasi:2026ltw}, or through simulation-based inference~\citep{Cranmer:2019eaq}.
The latter overcomes the computational bottleneck by training deep neural network architectures (such as normalising flows~\citep{Kobyzev:2019ydm,Papamakarios:2019fms}) on millions of simulated data examples to ``amortise'' the statistical information about gravitational wave parameters. 
After training, the networks can produce posterior samples of an event in seconds.
Simulation-based inference has widely been used in GW inference~\citep{Green:2020hst,Dax:2021tsq,Roulet:2026mzz}, particularly for low-latency PE of BBH signals~\citep{Chatterjee:2024pbj,Marx:2025ioo}, long-duration BNS signals in current detectors~\citep{Dax:2024mcn} or next-generation detectors~\citep{Hu:2024lrj, Alvey:2024uoc}.
However, this comes with the burden of having to train these networks, which can easily take on the order of days~\citep{Dax:2021tsq,Dax:2022pxd} and requires additional effort to achieve the same flexibility as already offered by stochastic samplers~\citep{Kofler:2025dux}.
Among these approaches, the \texttt{DINGO-BNS} pipeline~\citep{Dax:2024mcn} arguably represents the current state of the art for real-time PE, producing complete BNS posteriors in around one second once trained; we return to a qualitative comparison with this approach in \Cref{sec:discussion}.

In this work, we approach the bottleneck from a different perspective, driven by the large advances in graphics processing units (GPUs). Specifically, we extend our previous work~\citep{prathaban_gravitational-wave_2026} demonstrating the advantages of implementing nested sampling algorithms on a GPU for GW inference. For a given inference problem, vectorised nested sampling algorithms built on top of GPU-enabled waveform and detector modelling codes~\citep{Wong:2023lgb,Edwards2024ripple} enable massive parallelism of the parameter estimation task~\citep{Yallup:2025sty}. For long-duration signals, the degree to which this parallelism can be achieved is determined by the GPU memory, which for state-of-the-art chips is approaching $100$\,GB, enough to parallelise over the full frequency grid of a $128$\,s signal. This hardware-based fact allows us to build algorithms on the full frequency grid, without relying on any of the aforementioned likelihood acceleration techniques, relaxing the constraints on compression without additional cost. Beyond simple hardware acceleration, we also provide a complementary contribution that drives performance: a new nested sampling kernel designed specifically for gravitational wave inference. This Slice-within-Gibbs (SwiG) kernel~\citep{Yallup2026swig} is a GPU-native nested sampling outer loop, implemented within the \texttt{BlackJAX} ecosystem~\citep{Cabezas2024,jax2018}. Its structure is explicitly optimised for compact binary coalescence signals, with blocks chosen based on known modelling effects (specifically, that extrinsic parameters are much faster to sample by caching the generated GW waveform) and parameter degeneracies. The JAX-based waveform models that we utilise are developed within the \texttt{ripple} framework~\citep{Edwards2024ripple,Chan2026ripple}.

\subsection{Nested sampling and the constrained prior}

\noindent Nested sampling~\citep{Skilling2006,Ashton2022} is a Monte Carlo method that directly computes the Bayesian evidence $\mathcal{Z} = \int \mathcal{L}(\theta)\pi(\theta)\,\mathrm{d}\theta$, the integral of the likelihood $\mathcal{L}$ over a prior $\pi$ distribution over some parameters $\theta$. By maintaining a population of $m$ \emph{live points}, initially sampled from the prior, subsequently evolved to approximate the constrained prior,
\begin{equation}
\pi^\ast(\theta) \;\propto\; \pi(\theta)\,\mathbf{1}\!\left[\log\mathcal{L}(\theta) > \log\mathcal{L}^\ast\right],
\label{eq:cprior}
\end{equation}
following a constraint threshold $\log\mathcal{L}^\ast$ that is monotonically raised. At each iteration the $k$ live points with the lowest likelihood are removed (raising $\log\mathcal{L}^\ast$ to the next-worst live likelihood), their volume contributions are accumulated into the evidence estimator, and replaced by $k$ new draws from Eq.~\eqref{eq:cprior}. The dominant cost is sampling the constrained prior; the standard remedy in high dimension is constrained slice sampling, originating in PolyChord~\citep{Handley2015,Neal2003}, where each new live point is generated by a short MCMC chain that respects the hard constraint $\log\mathcal{L}>\log\mathcal{L}^\ast$. We use the vectorised JAX implementation of nested sampling from Ref.~\cite{Yallup2026}, which runs all $k$ replacement chains in parallel across particles.

\subsection{Slice-within-Gibbs}

\noindent When the parameter space splits into blocks $\theta = (\theta^{(1)},\ldots,\theta^{(B)})$, writing $\theta^{(-b)}$ for all blocks other than $b$, the constrained prior factorises as a sequence of \emph{block-conditional} constrained priors,
\begin{equation}
\begin{split}
&\pi^\ast\!\bigl(\theta^{(b)} \mid \theta^{(-b)}\bigr)
  \;\propto\; \pi\!\bigl(\theta^{(b)} \mid \theta^{(-b)}\bigr)\\
  &\quad\times\,\mathbf{1}\!\left[\log\mathcal{L}(\theta = (\theta^{(1)},\ldots, \theta^{(b)},\ldots,\theta^{(B)})) > \log\mathcal{L}^\ast\right],
\end{split}
\label{eq:block_cprior}
\end{equation}
and a Gibbs sweep over the $B$ blocks, each sampled by a constrained slice update, leaves $\pi^\ast$ invariant. This is the core idea of Nested SwiG~\citep{Yallup2026swig}, an outer nested sampling loop driven by an inner Slice-within-Gibbs kernel on the constrained prior. Such an update is an instance of \emph{Metropolis-within-Gibbs} (or coordinate-wise) sampling, a general-purpose family that leaves the full joint target invariant (up to MCMC convergence). Recent results establish dimension-robust convergence for Metropolis-within-Gibbs on broad classes of hierarchical models~\citep{ascolani2026scalability}, and show that it can be competitive with gradient-based samplers at scale~\citep{luu2025gibbs}. The slice primitive itself is the hit-and-run slice construction~\citep{smith1984efficient,Neal2003}, with proposal directions sampled from a block-diagonal covariance estimated from the live points, following Ref.~\cite{Handley2015}.

Some form of alternating block Gibbs sampling also appears in other astrophysical inference codes, especially when scaling inference to high dimensions; the LISA global fit~\citep{Littenberg:2020bxy} and the \texttt{Borg} model for initial matter conditions from large scale cosmology~\citep{BORG} give two prominent examples within the field. We differ primarily in exploiting this within nested sampling, and in applying these techniques to problems with less cleanly established decorrelation.

\subsection{What structure we exploit}

\noindent The setting that motivates Nested SwiG~\citep{Yallup2026swig} has a likelihood that factors over $J$ conditionally independent data groups, $\log\mathcal{L} = \sum_{j=1}^J \log\mathcal{L}_j(\theta_j,\psi)$, and the core algorithmic gain is the \emph{budget decomposition}: the global constraint $\log\mathcal{L}>\log\mathcal{L}^\ast$ rearranges into $J$ per-block budgets checkable in $\mathcal{O}(1)$, dropping the cost of a single Gibbs sweep from $\mathcal{O}(J^2)$ to $\mathcal{O}(J)$. A secondary, but equally important, gain is the improved \emph{mixing} the blocked Gibbs structure delivers. Each constrained slice update operates on a low-dimensional subspace where the constraint is locally simple, sidestepping the poor high-dimensional mixing of joint-space slice sampling. This mixing benefit is generic to the Gibbs decomposition itself, and does not require the data factorisation that powers the budget trick.

Compact-binary parameter estimation does not have the strict structure required to decompose the budget. The matched-filter likelihood of \Cref{sec:likelihood} is a single inner product over the frequency band, so the parameters are coupled in evaluation and every block proposal must still check the full likelihood against $\log\mathcal{L}^\ast$. There is, however, a clear hierarchy of parameter speeds between the intrinsic and extrinsic parameters of a CBC event. This mirrors the \emph{fast-slow}-style decompositions used in CMB cosmology codes~\citep{Lewis:2013hha,Handley2015}, which exploit a similar flavour of structure. By partitioning the parameters into expensive and cheap blocks with weakly coupled conditionals, the sampler gains its efficiency through cache reuse and improved mixing rather than through an $\mathcal{O}(1)$ constraint check. The standard CBC posterior is able to exploit both of these aspects:
\begin{enumerate}[label=(\roman*)]
\setlength{\itemsep}{2pt}
\item \emph{Cache reuse on the dominant cost.} The wall-clock cost of a likelihood evaluation for ground-based detectors is dominated by waveform generation~\citep{Wong:2023lgb,Edwards2024ripple}; detector projection and the analytic marginalisations of \Cref{sec:likelihood} are computationally cheaper in comparison. Splitting the parameters into a slow block (intrinsic, regenerates the waveform) and a fast block (extrinsic, reuses a cached whitened waveform) means a Gibbs sweep regenerates the waveform only on slow-block proposals, so extrinsic moves run at the cost of the projection alone.
\item \emph{Faster mixing under mild conditional dependence.} Although the likelihood couples all parameters in evaluation, the \emph{conditional} coupling between intrinsic and extrinsic blocks is mild. Slice-within-Gibbs operates on each of these sub-problems separately, so every constrained slice update sees a locally simpler geometry. Unblocked slice sampling on the equivalent joint space, by contrast, fails to mix at any practical inner-step budget (\Cref{fig:pp_nss}).
\end{enumerate}
The kernel of \Cref{sec:kernel} is therefore the SwiG construction re-blocked along the fast-slow axis of CBC inference rather than along a data-factorisation axis.

\section{Method}\label{sec:method}

\noindent This section introduces the SwiG construction of~\Cref{sec:background} for compact-binary inference. We specify the matched-filter likelihood and its analytic marginalisations (\Cref{sec:likelihood}), the fast-slow block structure and the constrained slice kernel that operates on it (\Cref{sec:kernel}), the multi-GPU sharding strategy (\Cref{sec:sharded}), and the heterodyned likelihood used for the headline real-time result (\Cref{sec:het}). The outer loop and slice primitives follow Ref.~\cite{Yallup2026}, implemented in BlackJAX~\citep{Cabezas2024} on top of JAX~\citep{jax2018}.

\subsection{Likelihood and marginalisations}\label{sec:likelihood}

\noindent For an event with $N_d$ detectors, frequency-domain strain data $d_i(f)$ and one-sided noise power spectral densities $S_{n,i}(f)$, the matched-filter log-likelihood is
\begin{equation}
\log\mathcal{L}(\theta)
  = -\frac{1}{2}\sum_{i=1}^{N_d}\langle d_i - h_i(\theta) | d_i - h_i(\theta)\rangle_i,
\label{eq:loglike}
\end{equation}
with $\langle a|b\rangle_i = 4\,\mathrm{Re}\!\int_{f_{\min}}^{f_{\max}} a^*(f)b(f)/S_{n,i}(f)\,\mathrm{d}f$. Templates $h_i(\theta)$ are generated from the \texttt{IMRPhenomPv2\_NRTidalv2} waveform evaluated by the differentiable JAX implementation in~\texttt{ripple}~\citep{Edwards2024ripple}, following the modular pipeline of Ref.~\cite{Wong:2023lgb}.

As is standard in GW PE~\citep{Veitch2015,Thrane2019,Singer:2015ema}, several parameters can be marginalised out on-the-fly rather than stochastically sampled, each through a closed-form or a one-dimensional reduction far cheaper than an MCMC update. The coalescence phase $\phi_c$ marginalises to $\log I_0\!\bigl(2|\langle d|h\rangle|\bigr)$ for the dominant $(2,\pm2)$ mode. The luminosity distance $d_L$, entering as the scaling $h\propto 1/d_L$, reduces to a log-sum-exp over a $1$D grid with prior $p(d_L)\propto d_L^{2}$. The coalescence time $t_c$ reduces to a log-sum-exp over a grid of time shifts, which can be computed by a single FFT on the uniform full-resolution grid, or by direct summation on the non-uniform heterodyned bins. On a GPU, using closed-form phase marginalisation is more for convenience. The phase could equally be marginalised by brute-force summation over a dense grid, exactly as we already do for time and distance, at negligible additional cost and with no measurable change in accuracy. Empirically, we find that attempting to directly sample $\phi_c$ leads to poor mixing for almost any kernel, as the waveforms we employ are not sensitive to this phase; it only enters as a knife-edge degeneracy in the likelihood. As a consequence, we always marginalise the phase, and would broadly recommend a similar approach for any stochastic sampler (noting that for waveforms with higher order modes, a grid marginalisation should remain valid).

For the remaining two parameters (time of coalescence $t_c$ and luminosity distance $d_L$) we explore the trade-off between sampling and marginalisation interchangeably. As both utilise grid marginalisation, posteriors for these can be reconstructed after the run at negligible cost (\Cref{app:gridmarg}). When performing the injection tests we marginalise the phase and distance, but keep the coalescence time as a sampled parameter in the Gibbs kernel. In practice, the time marginalisation is slightly more expensive than the distance marginalisation, as it sums the overlap over a grid of time shifts. At full frequency resolution either choice yields calibrated posteriors (\Cref{sec:pp}); under the heterodyned likelihood of \Cref{sec:het}, however, in our implementation we find that the time parameter is undercovered. We discuss this further in \Cref{sec:het} and \Cref{app:gridmarg}. We use phase and time marginalisation when recovering the GW170817 event, but sample the distance. The expensive cost is mostly concentrated in evaluating the waveform $h(f;\theta_{\mathrm{intr}})$ (where $\theta_{\mathrm{intr}}$ refer to the intrinsic parameters of the waveform model) and projecting it onto the detector network, both of which are dominated by the intrinsic parameter update.

\subsection{Blocked nested slice sampling kernel}\label{sec:kernel}

\noindent We sample over the remaining $d=15$ parameters: chirp mass $\mathcal{M}_c$, mass ratio $q$, six dimensionless component spins (sphere parameterisation, $\chi<0.05$), inclination $\iota$, two tidal deformabilities $\Lambda_1,\Lambda_2$, sky and polarisation $(\alpha,\delta,\psi)$, and either the coalescence time $t_c$ or distance $d_L$. The intrinsic parameters $\theta_s=\{\mathcal{M}_c,q,\mathbf{s}_1,\mathbf{s}_2,\iota,\Lambda_1,\Lambda_2\}$ require regenerating the waveform on every update and are termed \emph{slow}\footnote{ The inclination would be more naturally considered an extrinsic parameter; however, due to how it enters the waveform calculation in the \texttt{ripple} code we include it as \emph{slow}.}; the projection-only parameters $\theta_f=\{\alpha,\delta,\psi,t_c / d_L\}$ reuse a cached whitened polarisation pair $(h_+,h_\times)$ and are termed \emph{fast}. We utilise two reparameterisations that are standard in GW PE: the chirp-mass/mass-ratio parameterisation of the masses, and the detector-based sky frame (zenith and azimuth relative to a detector baseline) in which we sample the source direction rather than in equatorial $(\alpha,\delta)$~\citep{bilby_paper}. Both decorrelate parameters that are otherwise strongly coupled, complementing the block decomposition below. Additional standard reparameterisations --- for example, detector-frame time --- would likely help the blocked approach.

Unless otherwise specified each nested sampling replacement deletes the $k=64$ lowest-likelihood particles from a live set of $m=512$ and resamples them under the constraint $\log\mathcal{L}>\log\mathcal{L}^\ast$, where $\log\mathcal{L}^\ast$ is the next-worst live likelihood. The constrained inner step is a Gibbs sweep over seven sub-blocks of $(\theta_s,\theta_f)$:
\begin{itemize}
    \item \textbf{Slow blocks:} 
    $\{\mathcal{M}_c,q,\Lambda_1,\Lambda_2\}$, $\mathbf{s}_1$, $\mathbf{s}_2$, and $\iota$.
    \item \textbf{Fast blocks:} 
    $\{\alpha,\delta\}$ (sky-frame), $\psi$, and $\{t_c / d_L\}$.
\end{itemize}
The grouping within each sub-block follows the converse of the weak-coupling argument used to justify the Gibbs structure. Parameters that we expect to be \emph{strongly} conditionally correlated are kept together, so that joint hit-and-run moves can mix along the local degeneracy direction rather than fighting it. Chirp mass, mass ratio and the tidal deformabilities all enter the inspiral phasing at adjacent post-Newtonian orders and trade off against one another in a well-known mass--tidal degeneracy, so $\{\mathcal{M}_c,q,\Lambda_1,\Lambda_2\}$ are grouped. Each spin couples its three sphere-parameter components $\{|\mathbf{s}_i|,\theta_i,\phi_i\}$ tightly through the spin-to-Cartesian conversion, so $\mathbf{s}_1$ and $\mathbf{s}_2$ are kept as joint sub-blocks; the small spin bound $\chi_{\max}=0.05$ keeps the live points clear of the parameterisation's coordinate singularities and we observe no associated pathology. Sky position $(\alpha,\delta)$ is jointly constrained by inter-detector arrival-time triangulation, and is grouped accordingly. In this setting, the remaining parameters ($\iota$, $\psi$, $t_c$) couple to the rest of the space only weakly under the constraint and are updated as singletons. Within each sub-block a hit-and-run slice update~\citep{smith1984efficient,Neal2003} is applied along a random direction drawn from the block-diagonal covariance estimated from the live points, using the standard slice shrinking/expansion procedure~\cite{Handley2015}.

The slow blocks share a single waveform cache. At the start of each replacement we recompute $(h_+,h_\times)$ at the current intrinsic parameters, then pass the cache through the slow sub-blocks, recomputing and overwriting the cache on a new slow-block proposal; on the fast blocks the cache is fixed, so only the projection and the analytic marginalisations are re-evaluated.

The role of the Gibbs blocking is the same as in Nested SwiG~\citep{Yallup2026swig}; the constrained prior is hard for joint-space slice sampling because the extrinsic projection couples poorly to the intrinsic chirp parameters, and unblocked nested slice sampling at any reasonable inner-step budget fails to mix on the resulting manifold (\Cref{fig:pp_nss}). Blocking restricts each slice update to a subspace on which the constraint is locally simple and recovers calibrated posterior coverage at a much smaller compute budget. We expose the number of inner Gibbs sweeps per replacement, $M$, as a tunable parameter, defaulting to a single sweep, $M=1$. The outer loop is the standard NSS iteration, run to a termination threshold of $\log\mathcal{Z}_{\mathrm{live}} - \log\mathcal{Z}_{\mathrm{dead}} < -3$.

\subsection{Exploiting parallelism}\label{sec:sharded}

\noindent In this work, we exploit parallelism at two complementary levels: within-device parallelism and across-device parallelism. These provide complementary avenues for optimisation.

\vspace{8pt}
\noindent \emph{Within-device parallelism.} In practice, established physics nested sampling codes (\texttt{PolyChord}~\citep{Handley2015}, \texttt{dynesty}~\citep{2020MNRAS.493.3132S}, and similar) have historically favoured black-box kernels for ease of use, treating the sampler as a monolithic likelihood-constrained loop. We instead build on the composable transform system underlying BlackJAX~\citep{Cabezas2024}, which lets custom kernels such as ours be prototyped rapidly and, crucially, lets us vectorise what would otherwise be a sequential, particle-by-particle Gibbs sweep across the full batch of replacement chains, distributing the work over the many threads of a single GPU to saturate it and achieve high throughput. The deletion count of the Baseline configuration, $k=64$ (\Cref{sec:configs}), is chosen so that this per-replacement batch of constrained chains approximately saturates the target hardware for the uncompressed likelihood and chosen waveform; the optimal value should be re-evaluated for the available resources, compression level of the target problem and memory footprint of the waveform model.

\vspace{8pt}
\noindent \emph{Across-device parallelism.} Where the problem permits, it is equally useful to exploit the trivial parallelism available across multiple devices. For CPU-based inference this has traditionally been essential, with samplers distributed across many cores via MPI~\citep{2020MNRAS.498.4492S}. The analogous strategy on GPUs --- distributing across the devices of a multi-GPU node, such as the four GH200 superchips that make up an Isambard-AI node~\citep{mcintosh2024isambard} --- is a natural design target. The trade-off between compute and communication is different on a GPU. The in-node NVLink fabric coupling the GPUs is faster than the network underpinning CPU MPI in both bandwidth and latency, but the per-device compute is so fast that cross-device synchronisation quickly becomes the bottleneck. A GPU-distributed inference algorithm is therefore more sensitive to how often the devices must synchronise than to the total volume of data they exchange.

Vectorising the kernel across particles already saturates on-device memory for the 128\,s segment on a single GPU. To extend beyond one device we shard the live set across $D=4$ GPUs using JAX's \texttt{shard\_map}, partitioning the particles along the live axis while replicating the integrator state (the running $\log\mathcal{Z}$ estimator). Only two collective operations per replacement are then required:
\begin{enumerate}[label=(\roman*),nosep]
\item an all-gather of the $m$ live log-likelihoods, used to find the $k$ global worst particles and the survivor pool;
\item an all-gather of the full live particle states, used to sample $k/D$ starting points per device.
\end{enumerate}
Each device then runs $k/D=16$ constrained MCMC chains in parallel with the same Gibbs kernel described above, and both $m$ and $k$ are chosen divisible by $D$. The expensive likelihood and waveform work stays entirely local to each device, so a full outer iteration costs only these two all-gathers of communication.

\subsection{Likelihood compression}\label{sec:het}

\noindent For long-duration signals the vast number of frequency bins that are observed naturally motivates the use of some kind of data compression. For the compressed runs, we adopt a simple heterodyned (relative-binning) likelihood~\citep{Cornish:2010kf,Zackay:2018qdy,Cornish:2021lje,Krishna:2023bug}. The 128\,s segment at $f\in[20,2048]$\,Hz contains ${\sim}2.6\times10^{5}$ frequency bins per detector, of which only a tiny fraction carry information once a reference waveform $h_0$ near the peak likelihood is available. We use $N_b=5000$ bins of equal maximum phase change, following the implementation of Ref.~\cite{Wong:2023lgb}; the bin grid is set by the source phasing and shared across the network, while the per-detector summary data and the linearly-expanded waveform ratio $r_i(f)=h_i(f;\theta)/h_{0,i}(f)$ carry each detector's antenna projection, time delay, and PSD weighting. Evaluating the network likelihood, for $N_d=3$ detectors, then costs $N_b N_d = 1.5\times10^{4}$ operations rather than the ${\sim}7.8\times10^{5}$ of the full-resolution sum, independent of the underlying frequency resolution, and the phase, time and distance marginalisations act on the binned inner products unchanged. The reference waveform is taken at the injected parameters; we return in \Cref{sec:discussion} to locating it from a coarse pre-run for real events.

One parameter requires some care under this compression. The binning assumes the waveform ratio $r_i(f)$ varies smoothly across each bin, which holds for the intrinsic and most extrinsic parameters but is a poorer approximation for the coalescence time, whose effect is a rapidly winding phase that the coarse bins track only imperfectly. We find that sampling $t_c$ on a broad prior directly against the binned likelihood can fail the calibration test in $t_c$ while retaining good coverage in every other parameter (\Cref{app:gridmarg}). Whether this reflects a genuine limitation of heterodyning at this bin count or a detail of our implementation, it is easily sidestepped by marginalising the coalescence time in place of the distance, which we demonstrate in \Cref{app:gridmarg} restores coverage of all sampled parameters. We adopt this time-marginalised form for the GW170817 analysis (\Cref{sec:gw170817}); both marginalisation choices are validated in \Cref{app:gridmarg}.

This kind of compression also acts as a guide for the throughput achievable for shorter duration segments. Combined with fast-alert non-precessing waveforms, typical $4$\,s BBH segments can be analysed in comfortably less than a minute, even on commercial-grade hardware. We emphasise that the kernel is agnostic to the choice of compression: heterodyning is adopted here for its simple JAX implementation, but alternatives such as reduced order quadrature~\citep{Canizares:2014fya,Smith:2016qas,Morisaki:2023kuq} or multibanding~\citep{Garcia-Quiros:2020qlt,Morisaki:2021ngj} expose the same cached fast-slow likelihood structure required by \Cref{sec:kernel}, and could be substituted without modification to the sampler.

\section{Experiments}\label{sec:experiments}

\subsection{Injection catalogue}\label{sec:catalogue}

\noindent We validate the kernel on a catalogue of $N=1000$ synthetic BNS injections generated under the recovery prior, of which the first $100$ are used for the main-text P--P test and all $1000$ are used in the stress test of \Cref{app:stress}. Injection parameters are drawn independently per event from the priors in \Cref{tab:priors}, evaluated with \texttt{IMRPhenomPv2\_NRTidalv2} into the same H1, L1, V1 network used for recovery. Each signal is added to an independent realisation of coloured Gaussian noise drawn from the design-sensitivity PSDs (\texttt{aLIGO\_O4\_high} for H1/L1, \texttt{AdV} for V1, taken from \texttt{Bilby}~\citep{bilby_paper}); the noise seed is fixed per event so the catalogue is reproducible. We use $128$\,s of data centred $2$\,s before merger, $f\in[20,2048]$\,Hz, and a fixed reference GPS time of $1\,187\,008\,882$ (the GW170817 trigger). The noise realisation itself is independent of this time, but it sets the Earth orientation (and hence the antenna patterns and inter-detector delays) that projects a given sky position onto the H1, L1, V1 network; injection and recovery share it, so its specific value is an arbitrary but self-consistent convention. Each injection's catalogue entry stores the full set of parameters, the per-detector optimal SNRs, and per-event seeds for both the noise realisation and the sampler initialisation. Injection and recovery use identical priors in all $17$ parameters, so the catalogue is drawn from the same uniform spin-sphere prior ($\chi_{\max}=0.05$) that the sampler explores --- there are no zero-spin or aligned-spin injections. The network SNR distribution is right-skewed with median $24$ (mean $27$), ranging from $3.5$ to $140$ across the catalogue (\Cref{fig:snr}); GW170817, at network SNR ${\sim}32$, sits comfortably within the bulk of this distribution. The $d_L$ prior $[30,150]$\,Mpc is deliberately narrow and matches the injection distribution exactly, so any truncation of the posterior at the prior edges is consistent between injection and recovery and does not bias the P--P calibration.

\begin{table}
\caption{\label{tab:priors}%
Recovery (and injection) priors for the BNS catalogue. Spin sphere
parameterisation uses $|\mathbf{s}_i|\in[0,\chi_{\max}]$, polar angle
$\theta_i$ uniform in $\cos\theta_i$, and azimuth $\phi_i$ uniform in
$[0,2\pi)$. Phase $\phi_c$ and luminosity distance $d_L$ are analytically
marginalised (\Cref{sec:likelihood}); the $d_L$ range sets the
marginalisation grid, not a sampled dimension. All other parameters,
including $t_c$, are sampled by the kernel of \Cref{sec:kernel}.}
\begin{ruledtabular}
\begin{tabular}{lll}
Parameter & Range & Prior \\
\colrule
$\mathcal{M}_c\,[M_\odot]$       & $[1.5,\,2.5]$        & Uniform \\
$q$                              & $[0.5,\,1.0]$        & Uniform \\
$|\mathbf{s}_1|,|\mathbf{s}_2|$  & $[0,\,0.05]$         & Uniform \\
$\cos\theta_{s_i}$               & $[-1,\,1]$           & Uniform \\
$\phi_{s_i}$                     & $[0,\,2\pi)$         & Uniform \\
$\iota$                          & $[0,\,\pi]$          & $\sin\iota$ \\
$\Lambda_1,\Lambda_2$            & $[0,\,5000]$         & Uniform \\
$\alpha$                         & $[0,\,2\pi)$         & Uniform \\
$\delta$                         & $[-\pi/2,\,\pi/2]$   & $\cos\delta$ \\
$\psi$                           & $[0,\,\pi)$          & Uniform \\
$t_c\,[\mathrm{s}]$              & $[-0.1,\,0.1]$       & Uniform \\
$\phi_c$                         & $[0,\,2\pi)$         & Uniform (marg.) \\
$d_L\,[\mathrm{Mpc}]$            & $[30,\,150]$         & $\propto d_L^{2}$ (marg.) \\
\end{tabular}
\end{ruledtabular}
\end{table}

\subsection{Sampler configurations}\label{sec:configs}

\noindent We test four different configurations for the sampler.
All configurations share the catalogue, the waveform, the analytic marginalisations and the Gibbs block structure of \Cref{sec:kernel}; they differ only in the hyperparameters summarised in \Cref{tab:configs}. The \emph{Baseline} uses $m=512$ live points, deletes $k=64$ per outer iteration, performs one Gibbs sweep per replacement and evaluates the likelihood at full frequency resolution. The \emph{High-Res} variant runs $M=3$ sweeps to tighten posterior coverage at fixed live-point budget. The \emph{Sharded} variant matches Baseline but distributes the $m,k$ across $D=4$ GPUs as described in \Cref{sec:sharded}. The \emph{Heterodyned} variant replaces the full likelihood with the relative-binning likelihood of \Cref{sec:het} at $N_b=5000$ phase-spaced bins, keeping all other settings identical to Baseline. A time-marginalised counterpart, used for the GW170817 analysis, is described and validated in \Cref{app:gridmarg}. We also include an unblocked nested slice sampler (NSS) as a control for the role of the fast-slow block structure (\Cref{fig:pp_nss}). For NSS there are no blocks to sweep over; following the standard \texttt{PolyChord} convention~\citep{Handley2015} we instead set the number of joint slice steps per outer iteration to $M$ times the parameter-space dimension $d$, with $M=3$ chosen to match the per-iteration work of the High-Res configuration.

\begin{table}
\caption{\label{tab:configs}%
Sampler configurations. All variants use the kernel of
\Cref{sec:kernel} with grouped slow blocking; $m$ is the live-set size,
$k$ the number of deletions per outer iteration, $M$ the inner Gibbs
sweep count, and $D$ the number of GPUs. $^*$NSS is an unblocked
nested slice sampler used as a control in \Cref{fig:pp_nss}; having no blocks, it instead runs $M$ joint slice steps per parameter-space dimension $d$ (the standard \texttt{PolyChord} convention), with $M=3$ matching the work of the High-Res configuration.}
\begin{ruledtabular}
\begin{tabular}{lccccl}
Configuration & $m$ & $k$ & $M$ & $D$ & Likelihood \\
\colrule
Baseline       & $512$  & $64$    & $1$ & $1$ & full \\
High-Res       & $512$  & $64$    & $3$ & $1$ & full \\
Sharded        & $512$  & $64$    & $1$ & $4$ & full \\
Heterodyned    & $512$  & $64$    & $1$ & $1$ & relative bin. \\
NSS & $512$ & $64$ & $3^*$ & $1$ & full, unblocked \\
\end{tabular}
\end{ruledtabular}
\end{table}

\subsection{Hardware and software}\label{sec:hardware}

\noindent The pipeline is implemented on top of \texttt{Jim}~\citep{Wong:2023lgb} for the detector geometry and frequency-domain data interface, \texttt{ripple}~\citep{Edwards2024ripple} for differentiable JAX waveforms, and a custom matched-filter likelihood layer (\Cref{sec:likelihood}) that exposes the cached slow-likelihood / cached fast-likelihood pair required by the Gibbs kernel. The nested sampling outer loop and slice primitives extend BlackJAX~\citep{Cabezas2024}; the SwiG construction follows \Cref{sec:background,sec:kernel}. All timings are wall-clock on the Isambard-AI supercomputer~\citep{mcintosh2024isambard}. Each compute node hosts four NVIDIA GH200 Grace Hopper Superchips, where each superchip pairs a 72-core Grace CPU (120\,GB of memory) with an H100 Tensor Core GPU (96\,GB of memory), the four superchips within a node being connected by NVIDIA NVLink-C2C. We run the single-GPU configurations on one GH200 superchip, while the Sharded variant uses all four GH200s within a node; per-event times exclude one-off JIT compilation. Posteriors are processed with \texttt{anesthetic} and reweighting is performed using the nested sampling weights directly, with no resampling step.

\section{Results}\label{sec:results}

\subsection{Injection tests and P--P calibration}\label{sec:pp}

\noindent We validate the kernel on $N=100$ BNS injections drawn from the recovery prior, recovering with the same \texttt{IMRPhenomPv2\_NRTidalv2} waveform used for the injection (see \Cref{sec:method,sec:experiments}). For each injection and each parameter we compute the credible level of the true value, $q = \int w \, \mathbf{1}(\theta < \theta_{\mathrm{true}}) \, \mathrm{d}\theta$, using the nested sampling weights $w$ directly to avoid resampling artefacts. A well-calibrated posterior produces a uniform distribution of these credible levels; we summarise each run with per-parameter Kolmogorov--Smirnov $p$-values and a Fisher-combined $p$-value across all $15$ sampled parameters. Only the analytically marginalised parameters are excluded; their posteriors are reconstructed post-hoc by resampling the exact per-sample conditional on the marginalisation grid, so they are calibrated by construction, but the grid quantises the credible levels and invalidates the continuous null distribution of the KS statistic (\Cref{app:gridmarg}).

\begin{figure*}
  \centering
  \begin{subfigure}{0.49\textwidth}
    \includegraphics[width=\linewidth]{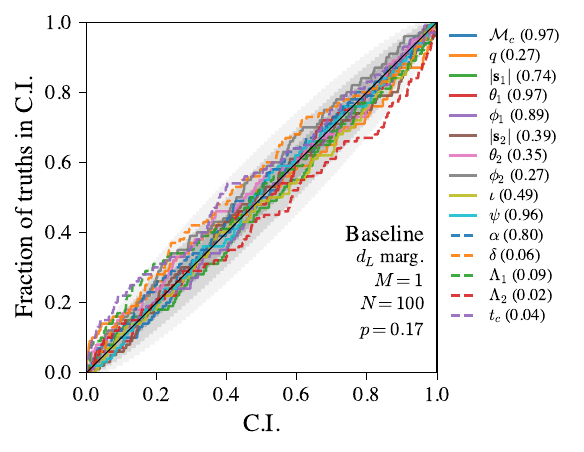}
    \caption{Baseline (full resolution, $M=1$).}\label{fig:pp_baseline}
  \end{subfigure}\hfill
  \begin{subfigure}{0.49\textwidth}
    \includegraphics[width=\linewidth]{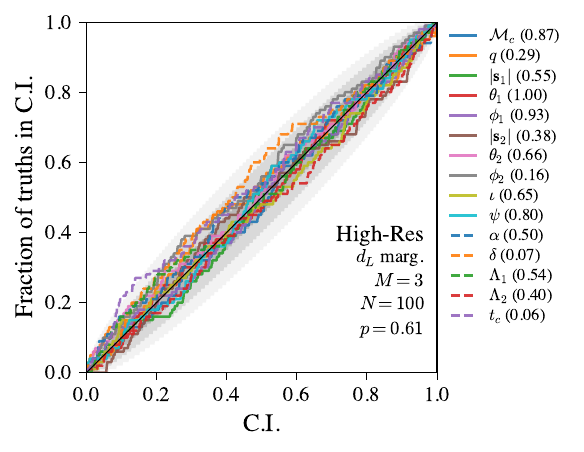}
    \caption{High-Res (full resolution, $M=3$).}\label{fig:pp_highres}
  \end{subfigure}\\[1.5ex]
  \begin{subfigure}{0.49\textwidth}
    \includegraphics[width=\linewidth]{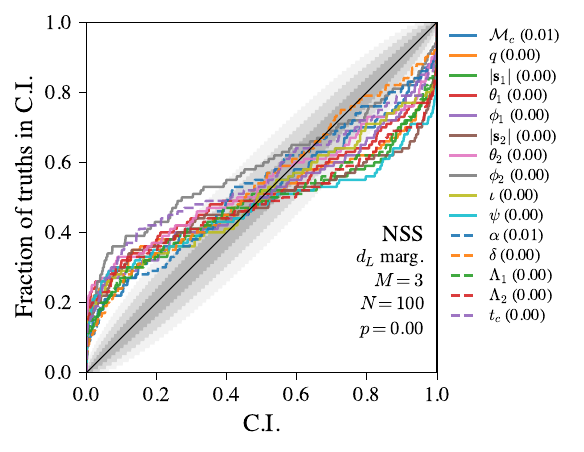}
    \caption{Unblocked NSS ($M=3$).}\label{fig:pp_nss}
  \end{subfigure}
  \caption{\label{fig:pp}%
    P--P diagnostics on the first $100$ injections, over all $15$ sampled
    parameters including $t_c$: (a) the out-of-the-box Baseline kernel,
    calibrated at a single Gibbs sweep per replacement; (b) the High-Res
    kernel ($M=3$), which produces the tightest agreement with the diagonal;
    and (c) the unblocked nested slice sampler, run with $M=3$ to match the
    work of High-Res, for which removing the fast-slow block structure leaves
    every parameter severely under-covered and collapses the Fisher-combined
    $p$-value to essentially zero. Shaded bands are the $1/2/3\sigma$ binomial
    confidence regions; legend entries give each parameter's KS $p$-value and
    the in-panel annotation the Fisher-combined $p$.}
\end{figure*}

\Cref{fig:pp} shows the P--P diagnostics for the three headline configurations. The out-of-the-box Baseline kernel (with a single Gibbs sweep) is calibrated across all $15$ sampled parameters with Fisher-combined $p = 0.17$ (\Cref{fig:pp_baseline}). Increasing the number of inner sweeps to three (High-Res, \Cref{fig:pp_highres}) raises the combined $p$ to $0.61$, tightening the agreement with the diagonal as the replacement chains decorrelate more fully --- a point we return to in \Cref{sec:blocking} --- at the cost of a longer wall time. Sharding the live set across four GPUs does not affect the calibration and produces consistent results with the Baseline, as expected for a faithful parallelisation. By contrast, removing the block structure altogether is catastrophic. The unblocked nested slice sampler (\Cref{fig:pp_nss}), run with $M=3$ slice steps to match the total work of High-Res, has a Fisher-combined $p$-value of essentially zero, with every parameter severely under-covered. This confirms that the fast-slow blocking of \Cref{sec:kernel} is what enables calibrated posteriors at these wall times; unblocked slice sampling in the full parameter space fails to mix at any reasonable computational budget.

The Heterodyned likelihood reproduces this calibration at a fraction of the cost in every parameter but the sampled coalescence time. This is confined to $t_c$; coverage of the physically relevant parameters remains valid with a combined $p = 0.15$, so the configuration is usable as-is. Marginalising $t_c$ instead, as in the GW170817 analysis of \Cref{sec:gw170817}, restores full calibration. Compressed inference is therefore exact once the coalescence time is handled on the grid rather than sampled, as detailed in \Cref{app:gridmarg}. Due to the success of the Baseline configuration, which otherwise shares all relevant nested sampling parameters, we conclude that the algorithm explores the space adequately, and that the miscalibration stems from a detail of our relative-binning implementation.

\begin{figure}
  \centering
  \includegraphics[width=\linewidth]{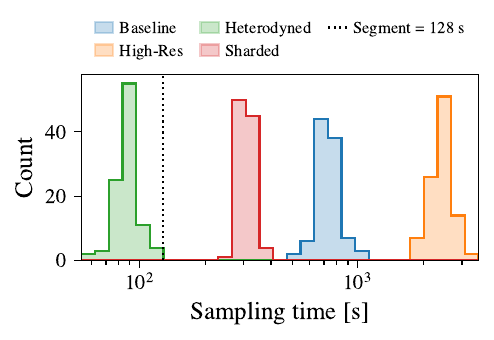}\\[1.5ex]
  \begin{ruledtabular}
  \begin{tabular}{lrrr}
  Configuration & Median [s] & Min [s] & Max [s] \\
  \colrule
  Baseline      &  726 &  512 & 1017 \\
  High-Res      & 2378 & 1775 & 3252 \\
  Sharded       &  306 &  258 &  383 \\
  Heterodyned   &   89 &   59 &  126 \\
  \end{tabular}
  \end{ruledtabular}
  \caption{\label{fig:runtime}%
    Per-event sampling wall time across the four principal configurations
    over the $100$ injections, shown both as the full distribution (histogram)
    and summarised by its median, minimum and maximum (table). Times exclude
    the one-off JIT compilation. The dotted line marks the $128$\,s
    data-segment duration; the Heterodyned configuration is the only one whose
    entire distribution, and hence worst case, falls below the segment
    length, i.e.\ true sub-segment, real-time inference.}
\end{figure}

\Cref{fig:runtime} quantifies the speed--accuracy trade-off. The Baseline delivers calibrated posteriors in a median of $\sim 12$\,min; High-Res triples this for the additional precision. Sharding reduces the Baseline time by a factor of $\sim 2.4$ to a median of $\sim 5$\,min, bringing wall time to within a small multiple of the segment length. The Heterodyned likelihood achieves fastest throughput, with a median wall time of $89$\,s, shorter than the $128$\,s of data. Marginalising the coalescence time in addition, the configuration used for GW170817 (\Cref{sec:gw170817}), costs an approximate $1.5\times$ premium, still keeping wall time on par with the segment duration.

All configurations yield posteriors with an effective sample size of ${\sim}4300$ (median ${\sim}4100$), set by the live-point budget and the compression of the nested sampling run rather than by the likelihood, and each run returns the log-evidence as a by-product. We do not probe the accuracy of the evidence in this work; however, both the evidence accuracy and the ESS of the posterior samples are straightforward to increase. Scaling the live-set and deletion counts together, $(m,k)\to\kappa(m,k)$, leaves the number of outer iterations essentially unchanged while raising the effective sample size by a factor of ${\sim}\kappa$~\citep{Yallup2026}. For the Heterodyned configuration in particular, where the per-likelihood cost is low enough that the device is far from saturated, this scaling is almost free, yielding $\kappa$ times the posterior samples at little additional wall time, so the most compressed run can deliver a substantially larger sample set while remaining within the sub-segment regime.

\subsection{Analysis of GW170817}\label{sec:gw170817}
\begin{figure*}
  \centering
  \includegraphics[width=\linewidth]{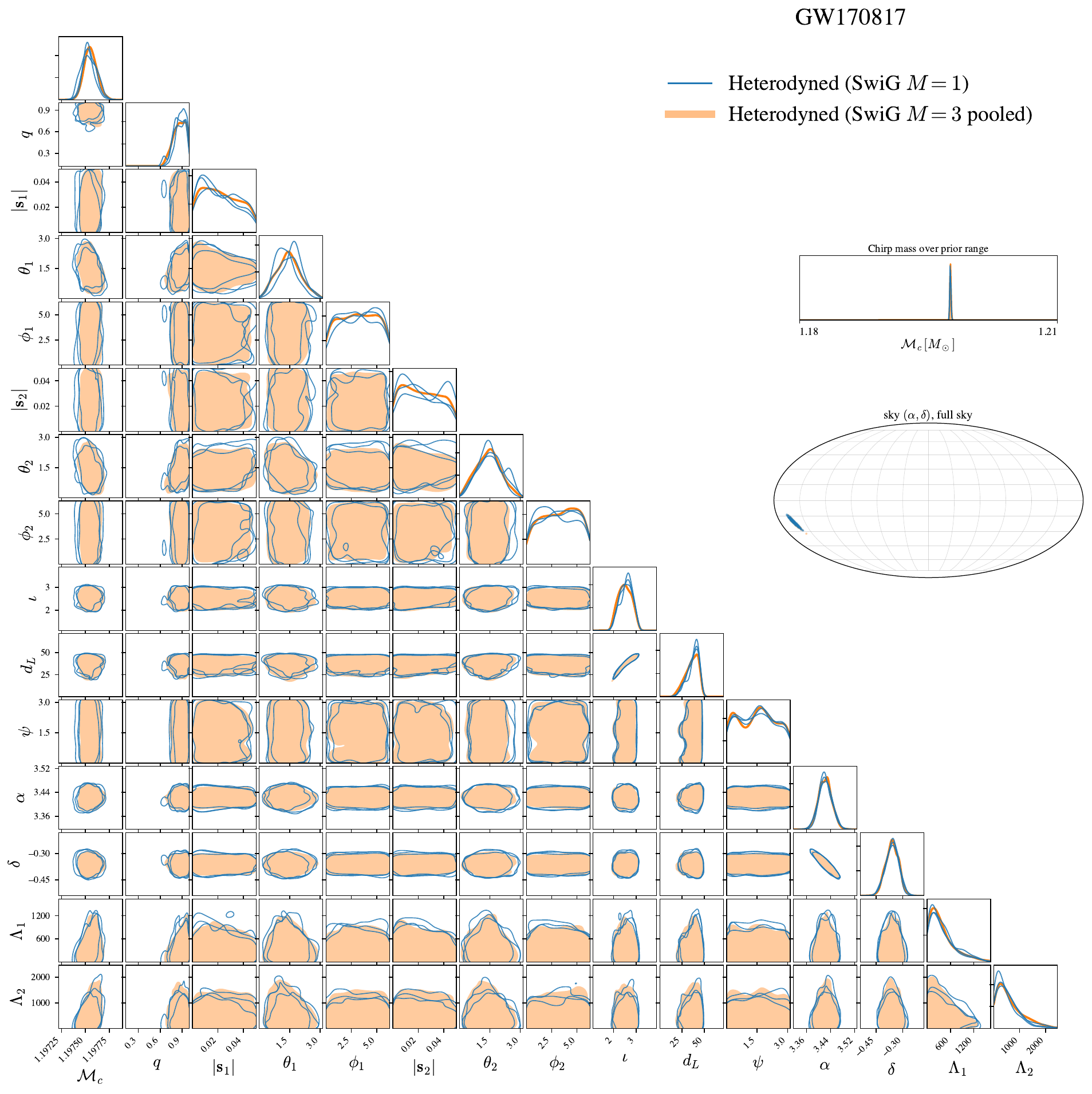}
  \caption{\label{fig:corner_event}%
    Posterior for the GW170817 event over all $15$ sampled parameters. Three independent Heterodyned (SwiG $M=1$) runs, each from a different random seed (blue $90\%$ contours), are overlaid on the pooled Heterodyned (SwiG $M=3$) reference obtained by combining three $M=3$ seeds (orange, filled $90\%$). The $M=1$ seeds are mutually consistent and reproduce the $M=3$ pooled reference across every parameter, demonstrating run-to-run reproducibility and convergence in the number of Gibbs sweeps.}
\end{figure*}

\noindent We apply the same configurations tested in~\Cref{sec:pp} to the real BNS event GW170817~\citep{LIGOScientific:2017vwq}. The data segment is $128$\,s long, centred on the trigger time GPS $1\,187\,008\,882$, and uses the same \texttt{IMRPhenomPv2\_NRTidalv2} waveform but swaps distance marginalisation for time marginalisation. Every parameter shown in \Cref{fig:corner_event} is sampled directly by the kernel, with no post-hoc reconstruction step. The Baseline, High-Res, Sharded and Heterodyned configurations are run with otherwise identical settings to those of \Cref{sec:pp}. In this case the reference-waveform parameters for the relative binning are found by maximum-likelihood optimisation, following the recipe of Ref.~\cite{Wouters:2024oxj}.

\Cref{fig:corner_event} shows the posterior recovered for the GW170817 event across all $15$ sampled parameters. The runs use the same priors as \Cref{tab:priors}, except for the chirp mass $\mathcal{M}_c\in[1.18,1.21]\,M_\odot$, the mass ratio $q\in[0.125,1]$, and the luminosity distance $d_L\in[1,75]\,\mathrm{Mpc}$, narrowed to ranges appropriate for the event. The coalescence-time marginalisation window is narrowed to $\pm0.03$\,s around the trigger, matching the \texttt{DINGO-BNS} setup~\citep{Dax:2024mcn} to enable the comparison of \Cref{sec:dingo_comparison}. As with the distance and phase marginalisations, this window sets only the extent of the marginalisation grid and not a sampled dimension: widening it back to the $\pm0.1$\,s of \Cref{tab:priors} enlarges the $t_c$ grid at modest additional cost (\Cref{app:gridmarg}). The cost grows linearly with the grid, however, so much wider windows eventually become prohibitive and are better handled by sampling $t_c$ directly. To demonstrate the run-to-run stability of the cheapest configuration we overlay $M=1$ runs from three consecutive random seeds against an $M=3$ reference formed by combining three $M=3$ seeds. We use a heterodyned likelihood, but include timing for full-resolution equivalents in \Cref{tab:gw170817_timings}. The single-seed $M=1$ contours are mutually consistent and reproduce the higher-resolution pooled $M=3$ posterior to within the resolution of the sample set across every parameter, confirming that a single inexpensive $M=1$ heterodyned run already recovers the converged posterior. Together with the P--P validation of \Cref{sec:pp}, this establishes that the heterodyned likelihood and the cheap $M=1$ sweep introduce no measurable bias on real data. As a complementary check, we rerun the injection configuration with the same time-marginalised likelihood and find this yields well-calibrated posteriors with a combined $p=0.51$ (\Cref{app:gridmarg}).

\Cref{tab:gw170817_timings} reports the timing breakdown for the four algorithms on GW170817. The sampling time follows the per-event budget of \Cref{fig:runtime}; the two one-off compilation costs --- the likelihood and the sampler-kernel JIT --- are paid once per run and are excluded from the quoted sampling times. We additionally list the number of waveform evaluations $N_{\mathrm{wf}}$ required to reach convergence, which is set by the number of Gibbs sweeps and the particle count. The High-Res configuration ($M=3$ sweeps) converges in roughly an hour, without any data compression and using broad, uninformative priors. The $M=1$ configs require approximately $1\times10^6$ waveform evaluations ($\sim\!4\times10^6$ likelihood evaluations once the cached fast updates are included), scaling to $3\times10^6$ waveforms ($\sim\!19\times10^6$ likelihood) for the $M=3$ configs. This is a significant efficiency gain in terms of waveform evaluations, and ensures that the algorithmic gains are compounded in situations where individual waveform evaluations become increasingly expensive.

\begin{table}
\caption{\label{tab:gw170817_timings}%
Timings for the four algorithms on the real event GW170817 with the
time-marginalised likelihood: post-JIT sampling time, one-off JIT compilation
(likelihood\,$+$\,sampler-kernel), and the approximate waveform evaluation count
($N_{\mathrm{wf}}$, set by the number of Gibbs sweeps $M$). Values are averaged
over three random-seed repeats. The Sharded sampler-kernel JIT is higher than
the Baseline owing to the multi-GPU compilation.}
\begin{ruledtabular}
\begin{tabular}{lrcc}
Configuration & Sampling [s] & JIT [s] & $N_{\mathrm{wf}}$ \\
\colrule
Baseline    &        1090 & $15 + 164$ & $10^6$ \\
High-Res    &        3766 & $11 + 221$ & $3\times10^6$ \\
Sharded     &         450 & $15 + 172$ & $10^6$ \\
Heterodyned &         125 & $7 + 164$  & $10^6$ \\
\end{tabular}
\end{ruledtabular}
\end{table}

\section{Discussion}\label{sec:discussion}

\subsection{Comparison with CPU-based stochastic sampling}\label{sec:cpu_comparison}

\noindent A closely matched configuration of the same event is tested in the \texttt{Bilby} validation paper~\citep{Romero-Shaw:2020owr}, using the \texttt{pBilby}~\citep{2020MNRAS.498.4492S} wrapper to distribute the \texttt{dynesty}~\citep{2020MNRAS.493.3132S} nested sampler across 560 CPU cores. The reported wall time with this setup is approximately $11$\,h. The comparison is not perfectly like-for-like: the \texttt{Bilby} analysis additionally samples detector calibration uncertainties (which we omit; \Cref{sec:limitations}), but conditions on the known sky position of the optical counterpart, where we sample the full sky, two differences that act in opposite directions on the difficulty of the inference. Setting aside the overhead of compile time (and finding a maximum-likelihood reference for the heterodyned likelihood), the Heterodyned configuration of this work produces a posterior we would expect to match in only $\sim\!125$\,s, while the sampler can converge on the full uncompressed data in $450$--$1090$\,s depending on resources. This represents a direct comparison speedup of up to a factor of $\sim\!100$ without any compression, and $\sim\!300$ when compressing the data.

The unblocked NSS control of \Cref{fig:pp_nss} helps demonstrate the source of this gain. Joint-space slice sampling is a reasonable structural proxy for the adaptive random-walk (``acceptance-walk'') kernels that drive production \texttt{dynesty} analyses~\citep{2020MNRAS.493.3132S,bilby_paper}: both evolve all parameters jointly through a sequence of likelihood-constrained moves, differing in the proposal primitive rather than in their treatment of the parameter space. The production kernels preserve calibration by adaptively growing the chain length, in practice expending orders of magnitude more likelihood evaluations per replacement than the fixed budgets used here~\citep{prathaban_gravitational-wave_2026}. We would expect an acceptance-walk kernel constrained to our matched per-iteration budget to fail in much the same way as NSS, and once run at its native adaptive budget it would only recover calibration by surrendering the wall-time gains. This is further exacerbated on GPU hardware as adaptive walk lengths lead to underutilising the throughput potential. The comparison between \Cref{fig:pp_highres} and \Cref{fig:pp_nss} therefore isolates the fast-slow blocking, rather than raw hardware throughput, as the ingredient that makes calibrated inference possible at these budgets.

\subsection{Comparison with amortised inference}\label{sec:dingo_comparison}

\noindent The \texttt{DINGO-BNS} pipeline~\citep{Dax:2024mcn} uses the same waveform, similar data-compression techniques and comparable GPU hardware, so serves as the natural comparison point for this work. The two approaches sit at opposite ends of the amortisation spectrum, and their strengths are largely complementary. \texttt{DINGO-BNS} invests days of GPU time~\citep{Dax:2021tsq,Dax:2022pxd} in training an expressive flow-based model of the posterior, tailored to a chosen prior and waveform model, as well as detector configuration and frequency ranges, unless masking is performed, e.g., with transformers~\citep{Kofler:2025dux}. 
Once trained, it produces posterior samples for any event matching that configuration in around one second, with importance sampling against the exact likelihood supplying a per-event accuracy diagnostic. The SwiG sampler is instead \emph{ab initio}: every analysis is launched cold from the prior with no offline stage, so a change of waveform model, prior range, noise spectrum, detector network or segment length carries no retraining cost. This freedom extends to the data compression itself. The heterodyning we adopt requires only a reference waveform, located per event, whereas reduced order quadrature --- the compression underpinning most recent low-latency LVK analyses~\citep{Chaudhary:2023vec,Morisaki:2023kuq} --- relies on a reduced basis built offline over the parameter space, a configuration-specific upfront cost of the same character as pretraining an amortised network. The same run returns the Bayesian evidence alongside the posterior, and its calibration is established directly by large-scale injection campaigns (\Cref{app:stress}). At one to two minutes per event this remains slower than a trained network, but it is comfortably within the latency at which alert information is consumed by observers, and it provides the robust, assumption-light complement that an amortised pipeline requires whenever the analysis configuration steps outside the trained one.

In the interest of a fair comparison, we note that the configuration differences bearing on it are, if anything, conservative. Our analyses retain the full $[20,2048]$\,Hz band --- roughly twice the frequency content of the \texttt{DINGO-BNS} configuration --- so adopting a matched band would roughly halve the cost of every full-resolution likelihood evaluation, and permit a proportionally coarser relative binning, translating directly into throughput. Several of the accelerations developed for \texttt{DINGO-BNS} --- notably event-adapted multibanded frequency grids and priors conditioned on information from the detection pipeline --- are likewise sampler-agnostic, and nothing prevents their adoption here; we expect them to compound with the gains already demonstrated. For completeness, the coalescence-time window of our heterodyned GW170817 analysis is matched to the narrow \texttt{DINGO-BNS} setting (\Cref{sec:gw170817}); because $t_c$ is marginalised there, its width has negligible bearing on cost either way. The two paradigms also compose naturally. For example, in situations where full precision is hard to amortise (e.g. very high SNR scenarios), SwiG nested sampling runs could be seeded by simulation-based inference proposals, refining the neural sampling results. 

\subsection{Limitations}\label{sec:limitations}

\noindent The fastest inference results in this work come with two caveats: the compilation time and the reference waveform. The former is purely a technical obstacle, so we do not consider it as part of the inference time. The latter is a more intricate issue. We do not find the reference parameters on the fly, so a realistic deployment would require a numerical routine to locate them. Prior work has established that reference-waveform parameters can be found in negligible time~\citep{Dax:2024mcn}, and that JAX-based optimization algorithms (compatible with our PE setup) can recover template parameters to seed the reference waveform in less than a minute~\citep{Green:2024yka}. 
We present results compressed to $5000$ bins per detector, a conservative choice that reduces the sensitivity to the reference waveform.

Two physics simplifications additionally separate our configuration from a full production setup. Contrary to the setup used in the LVK collaboration~\citep{Chaudhary:2023vec}, we do not include uncertainties in the detector calibration; as discussed in \Cref{sec:blocking}, calibration parameters sit naturally in the fast parameter block, and we expect their inclusion to be computationally cheap. Moreover, we do not include higher-order waveform modes, which are subdominant for the near-equal-mass BNS systems considered here, but have been shown to be important for systems with more unequal masses, such as NSBH mergers, where they carry significant information for low-latency follow-up~\citep{Iacovelli:2026xjj}. Their inclusion raises the cost of the slow (waveform) block but leaves the kernel structure of \Cref{sec:kernel} untouched.

\subsection{Choice of blocking scheme}\label{sec:blocking}

\noindent The choice of blocking presented in \Cref{sec:kernel} was informed primarily by what empirically performed well on the problem as defined by \Cref{tab:priors}. Some of the chosen structure may not generalise to all problems in the GW phase space; however, we believe there are significant lessons that can be transposed from this work. An example choice that may not generalise is the choice to keep the spins uncoupled and in separate blocks, which we would not expect to generalise well to BBH analysis where spins can have a more pronounced effect and should be coupled. A second, subtler sign that the blocking is effective but not optimal is the sweep count itself. The $M=1$ Baseline already produces calibrated posteriors (combined $p=0.17$), but a third sweep raises this to $p=0.61$ (\Cref{sec:pp}). A perfectly decorrelating block structure would leave nothing for the extra sweeps to do; the improvement instead reveals a weak residual conditional correlation between blocks that a single sweep does not fully break. The effect is not damaging to the conclusions, the single-sweep configuration passes comfortably, and we adopt it as our default. However, it indicates that the grouping of \Cref{sec:kernel}, while sufficient, is not the last word, and that a better-adapted decomposition might recover the same coverage in a single sweep. Whilst choosing a blocking involves some manual experimentation on the class of problems of interest, we propose that the massive increase in performance between the NSS and equivalent SwiG configuration demonstrates the potential payoff, as well as motivating investigation into tools that can automatically uncover this structure~\citep{hoffman_autoconj_2018}.

More broadly, the within-Gibbs structure exploited here is not specific to gravitational waves. While Slice-within-Gibbs is conceptually well established (\Cref{sec:background}), to our knowledge it remains unexploited in the large-scale nested sampling codes used across cosmology and astrophysics, where joint-space slice sampling~\citep{Handley2015} is standard. Many nested sampling problems at scale separate naturally into weakly coupled blocks with a single dominant expensive subspace, and would stand to benefit from the same caching and mixing gains demonstrated here. The detector calibration uncertainty previously mentioned is a concrete example within gravitational wave PE~\citep{Romero-Shaw:2020owr}. Calibration parameters sit firmly on the fast side of the fast-slow divide, re-weighting an already-generated waveform rather than regenerating it, and when grouped per detector form separate low-dimensional subspaces on which a blocked kernel mixes efficiently. They are equally a candidate for approximate marginalisation. Recent Laplace-based collapsed-sampling schemes~\citep{margossian2020hamiltonianmontecarlousing,lovick2026automatic} could remove this subspace from the sampled space entirely, in the same spirit as the phase and distance marginalisations of \Cref{sec:likelihood}. We leave a full treatment of detector calibration to future work.

\section{Conclusions}\label{sec:conclusions}

\noindent In this work, we have developed a new, GPU-native nested sampling algorithm, optimised for gravitational waves, that brings ab initio stochastic sampling into the real-time regime for binary neutron star signals observed with the current generation of GW detectors. At the core of this advancement is our bespoke Slice-within-Gibbs (SwiG) kernel, which explicitly leverages the well-known structural hierarchy of compact binary coalescence parameters. When combined with state-of-the-art GPU hardware, this allows us to establish four core results.
\begin{itemize}
\item \emph{Scalable, uncompressed inference.} By exploiting the full memory capacity of modern GPUs alongside the SwiG kernel, we achieve fully calibrated parameter estimation on the uncompressed frequency grid of a $128$\,s BNS signal in a median of twelve minutes on a single GPU. We validate this with extensive full P--P calibration tests with up to $1000$ binary neutron star injections (\Cref{fig:pp_gibbs3_1k}).

\item \emph{Sub-segment, real-time performance.} Mild data compression via relative binning drops the sampling wall time to a median of $89$ seconds, strictly shorter than the duration of the data segment itself, for every injection in the campaign. Analysing GW170817 with time marginalisation completes in around two minutes, and multi-GPU sharding closes much of the same gap ($\sim$5 minutes) with no data compression to the likelihood at all.

\item \emph{High-fidelity validation on GW170817.} Applying this framework to GW170817 recovers posteriors we would expect to match those of community-standard pipelines, converging in roughly two minutes rather --- a speedup of up to two orders of magnitude depending on configuration --- and, at the same time, returning the Bayesian evidence for model comparison.

\item \emph{The necessity of domain knowledge.} Our ablation studies explicitly confirm that unblocked nested slice sampling fails to mix at these tight computational budgets. Injecting the physical conditional independence and degeneracy structures of gravitational waves into the sampler geometry is the essential ingredient that makes inference at this speed possible, a lesson we expect to transfer well beyond the present application.
\end{itemize}
Taken together, these results establish a likelihood-based, training-free approach that is highly complementary to modern machine learning methods (such as amortised inference via \texttt{DINGO}), achieving comparable low-latency performance with classical sampling algorithms launched cold from the prior.

Looking further ahead, rapid parameter estimation for long-duration gravitational wave signals will only grow in importance. As we approach the era of next-generation ground-based observatories such as the Einstein Telescope and Cosmic Explorer, BNS signals will remain in-band for hours or even days, and even BBH signals will constitute minutes of data. Without optimisation and acceleration, extracting the full scientific output of these facilities will present a formidable computational challenge. By demonstrating that highly parallelised stochastic sampling, together with domain knowledge, can meet the extreme computational demands of modern long-duration signals, this work lays a robust, scalable, and exact foundation for next-generation GW science.

\section*{Acknowledgements}

\noindent The authors acknowledge the use of resources provided by the Isambard-AI National AI Research Resource (AIRR). Isambard-AI is operated by the University of Bristol and is funded by the UK Government’s Department for Science, Innovation and Technology (DSIT) via UK Research and Innovation; and the Science and Technology Facilities Council [ST/AIRR/I-A-I/1023]. Supported by the grant award ``Real Time Gravitational Wave Inference'' 0251-3025-2136-1. DY and WH are supported by the UKRI Frontier Research Guarantee [EP/X035344/1]. JA and NS are supported by fellowships from the Kavli Foundation. TCKN acknowledges support by the research program of the Netherlands Organization for Scientific Research (NWO). TW is supported by the research program of the Netherlands Organization for Scientific Research (NWO) through grant number OCENW.XL21.XL21.038. MP was supported by the Harding Distinguished Postgraduate Scholars Programme (HDPSP). DY, MP and WH were supported by the research environment and infrastructure of the Handley Lab at the University of Cambridge. The code developed for this study will be released as part of the \texttt{Jim} package~\citep{Wong:2023lgb}. We thank the GW~JAX~Team\footnote{\url{https://github.com/GW-JAX-Team}} for their continued maintenance of \texttt{Jim} and \texttt{ripple}.

This research has made use of data or software obtained from the Gravitational Wave Open Science Center~\citep{LIGOScientific:2019lzm}, a service of the LIGO Scientific Collaboration, the Virgo Collaboration, and KAGRA. This material is based upon work supported by NSF's LIGO Laboratory which is a major facility fully funded by the National Science Foundation, as well as the Science and Technology Facilities Council (STFC) of the United Kingdom, the Max-Planck-Society (MPS), and the State of Niedersachsen/Germany for support of the construction of Advanced LIGO and construction and operation of the GEO600 detector. Additional support for Advanced LIGO was provided by the Australian Research Council. Virgo is funded, through the European Gravitational Observatory (EGO), by the French Centre National de Recherche Scientifique (CNRS), the Italian Istituto Nazionale di Fisica Nucleare (INFN) and the Dutch Nikhef, with contributions by institutions from Belgium, Germany, Greece, Hungary, Ireland, Japan, Monaco, Poland, Portugal, Spain. KAGRA is supported by Ministry of Education, Culture, Sports, Science and Technology (MEXT), Japan Society for the Promotion of Science (JSPS) in Japan; National Research Foundation (NRF) and Ministry of Science and ICT (MSIT) in Korea; Academia Sinica (AS) and National Science and Technology Council (NSTC) in Taiwan.

\bibliography{main}
\appendix
\crefalias{section}{appendix}

\section{Grid marginalisations and post-hoc reconstruction}\label{app:gridmarg}

\noindent The analytic marginalisations of \Cref{sec:likelihood} share a single construction. Each of the three parameters $\{\phi_c, t_c, d_L\}$ enters the likelihood only through a cheap scalar operation on cached inner products, so the marginal likelihood over any subset of them can be evaluated by quadrature on a fixed grid. A Bessel function for $\phi_c$, a grid sum over time shifts for $t_c$, and a log-sum-exp over a one-dimensional distance grid for $d_L$, all at negligible cost relative to a waveform evaluation. The injection campaign of \Cref{sec:catalogue} marginalises $\{\phi_c, d_L\}$ and samples $t_c$, whereas the GW170817 analysis of \Cref{sec:gw170817} marginalises $\{\phi_c, t_c\}$ and samples $d_L$. Because a marginalised parameter is removed from the sampled space entirely, the evidence and the joint posterior of the remaining parameters are immune to any mixing pathology in that direction and the marginal is computed exactly, up to grid resolution, at every likelihood call.

Posterior samples for a marginalised parameter are reconstructed after the run. For each chosen point, the inner products defining the conditional posterior of the marginalised parameter on its grid can be evaluated, and a value drawn from this conditional directly. The reconstruction is pure resampling so it is calibrated by construction; the grid does, however, quantise the credible levels, invalidating the continuous null distribution of the KS statistic, which is why reconstructed parameters are excluded from the P--P tests of \Cref{sec:pp}. The cost scales with the number of samples reconstructed, and since most dead points carry negligible posterior weight, it suffices to reconstruct the retained posterior samples. At ${\sim}1$\,s per $1000$ samples this amounts to a few seconds for the ${\sim}4000$ effective samples of a typical run, and even this could be eliminated entirely by caching the inner products from the sampling phase. A configuration that samples a parameter directly pays no reconstruction cost for it.

The choice of which parameters to sample and which to marginalise interacts with the likelihood treatment. At full frequency resolution, sampling $t_c$ is unproblematic: the Baseline, High-Res and Sharded configurations are calibrated across all $15$ sampled parameters, including $t_c$ (\Cref{fig:pp,fig:pp_gibbs3_1k}). Under the heterodyned likelihood this no longer holds: as anticipated in \Cref{sec:het}, sampling $t_c$ against the binned surface can leave it miscalibrated. The effect is confined to a single parameter: the distance-marginalised Heterodyned configuration is calibrated in every parameter except $t_c$, whose lone KS $p$-value of $4\times10^{-5}$ collapses the Fisher-combined $p$ from $0.15$ (with $t_c$ excluded) to $0.003$. Whether this is intrinsic to relative binning at $N_b=5000$ or an artefact of our implementation, the physically relevant parameters retain valid coverage throughout, and the full-resolution results are unaffected.

\begin{figure}
  \centering
  \includegraphics[width=\linewidth]{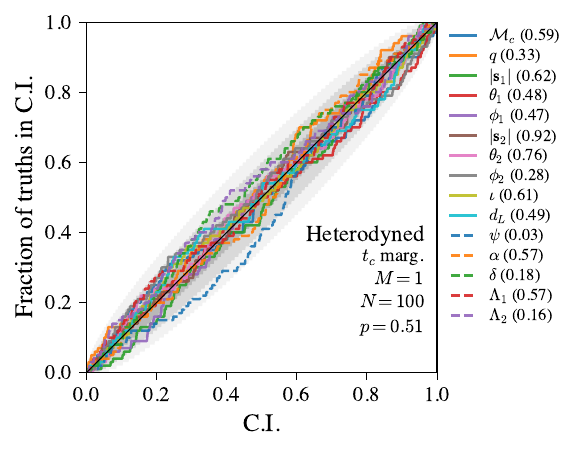}
  \caption{\label{fig:pp_het_tmarg}%
    P--P diagnostic for the time-marginalised heterodyned configuration
    ($t_c$ marginalised, $d_L$ sampled) on $100$ injections, with bands and
    legends as in \Cref{fig:pp}. All $15$ sampled parameters, including the
    directly sampled $d_L$ (KS $p = 0.49$), are calibrated, with
    Fisher-combined $p = 0.51$.}
\end{figure}

Marginalising the coalescence time instead, as in the time-marginalised heterodyned configuration of \Cref{fig:pp_het_tmarg}, integrates the time-shift factor essentially exactly on a dense grid and removes the effect entirely. The combined $p$ rises to $0.51$, and the sampled distance that takes $t_c$'s place is cleanly calibrated as a directly tested parameter (KS $p = 0.49$). This is the configuration adopted for the GW170817 analysis of \Cref{sec:gw170817}, at an approximate $1.5\times$ premium in sampling time over the distance-marginalised variant (median $132$\,s versus $89$\,s; \Cref{sec:pp}). Every parameter shown in \Cref{fig:corner_event} is therefore sampled directly by the kernel, and generally we recommend this as the baseline configuration.

\section{Stress-test at $N=1000$}\label{app:stress}

\begin{figure}[htb]
  \centering
  \includegraphics[width=\linewidth]{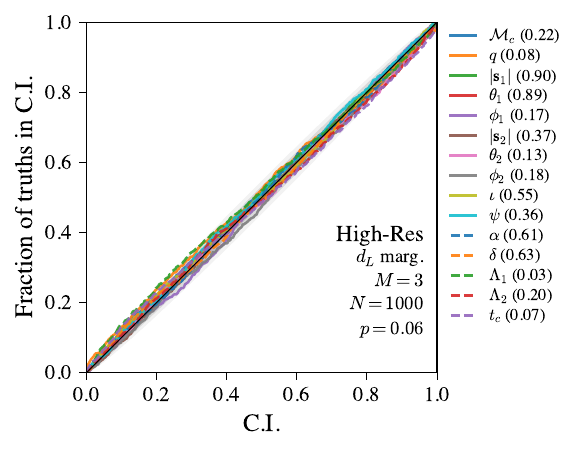}
  \caption{\label{fig:pp_gibbs3_1k}%
    P--P diagnostic for the High-Res configuration over the full catalogue
    of $N=1000$ injections, over all $15$ sampled parameters (excluding only
    the reconstructed $d_L$; \Cref{app:gridmarg}). All parameters lie within
    the binomial confidence bands and the Fisher-combined $p$-value is
    $0.06$.}
\end{figure}

\noindent The P--P validation in the main text uses $N=100$ injections, the standard budget in the GW literature. It is worth being precise about what passing such a test certifies. Under perfect calibration the credible levels are uniform whatever the catalogue size, so the $p$-value threshold is no harder to clear at large $N$; what grows with $N$ is the \emph{power} of the test --- the size of the miscalibration a pass rules out. The resolvable deviations are set by the binomial band $\sqrt{p(1-p)/N}$, which at $N=100$ has a $1\sigma$ width of $\sim 4.7\%$. A parameter systematically under-covered by $5\%$ at the $1\sigma$ level therefore produces only a $\sim 1\sigma$ excursion and passes far more often than not: a pass at $N=100$ certifies only the absence of biases at or above the $\sim 5\%$ level.

Increasing the catalogue size tightens this certificate as $1/\sqrt{N}$. At $N=1000$ the band narrows to $\sim 1.5\%$, and the same $5\%$ systematic, invisible at $N=100$, becomes a $\sim 3.5\sigma$ excursion that would collapse the combined $p$-value outright. \Cref{fig:pp_gibbs3_1k} shows the High-Res configuration applied to the full catalogue of $1000$ injections, over all $15$ sampled parameters including $t_c$. The combined $p$-value of $0.06$ passes this far more powerful test with no parameter falling outside its binomial confidence band. A pass at this budget certifies that any residual miscalibration is confined to the percent level --- a factor $\sqrt{10} \sim 3$ tighter than the main-text test can establish.

\begin{figure}[htb]
  \centering
  \includegraphics[width=\linewidth]{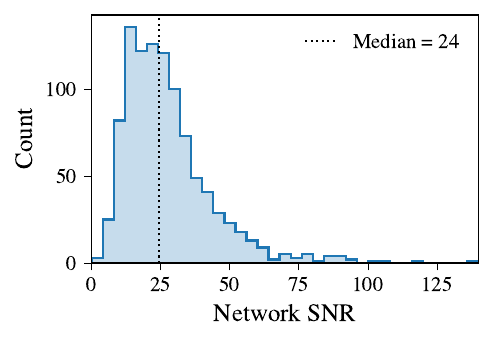}
  \caption{\label{fig:snr}%
    Network SNR distribution of the full $N=1000$ injection catalogue
    (median $24$, mean $27$, range $3.5$--$140$). GW170817, at network SNR
    ${\sim}32$, sits within the bulk of the distribution.}
\end{figure}

\Cref{fig:snr} shows the network SNR distribution of the full catalogue underlying this test, whose injection configuration is adapted from Ref.~\cite{Wouters:2024oxj}: right-skewed, with median $24$ and a loud tail extending to $140$, so the large-$N$ test also probes the high-SNR regime where posteriors are tightest and miscalibration is most easily exposed. GW170817, at network SNR ${\sim}32$, sits within the bulk of the distribution.
\end{document}